\documentclass[11pt]{article}
\usepackage[T1]{fontenc}
\usepackage[latin9]{inputenc}
\usepackage[a4paper]{geometry}
\usepackage[active]{srcltx}
\usepackage{textcomp}
\usepackage{mathrsfs}
\usepackage{amsmath}
\usepackage{amssymb}
\usepackage{esint}

\makeatletter
\usepackage{textcomp}

\pdfoutput=1 % if your are submitting a pdflatex (i.e. if you have
\usepackage{jheppub}% for details on the use of the package, please
\usepackage{etoolbox}% http://ctan.org/pkg/etoolbox
    
    \patchcmd{\maketitle}{\@fpheader}{}{}{}

\usepackage{amsfonts}

\usepackage{bbm}

\title{\boldmath Boundary conditions for General Relativity on AdS$_{3}$ and the KdV hierarchy}

\author[]{Alfredo P\'{e}rez,}
\author[]{David Tempo,}
\author[]{Ricardo Troncoso}

\affiliation[]{Centro de Estudios Cient\'{i}ficos (CECs), Av. Arturo Prat 514, Valdivia,
Chile.}

\emailAdd{aperez@cecs.cl}
\emailAdd{tempo@cecs.cl}
\emailAdd{troncoso@cecs.cl}

\preprint{CECS-PHY-16/04}

\abstract{It is shown that General Relativity with negative cosmological constant
in three spacetime dimensions admits a new family of boundary conditions
being labeled by a nonnegative integer $k$. Gravitational excitations
are then described by ``boundary gravitons'' that fulfill the equations
of the $k$-th element of the KdV hierarchy. In particular, $k=0$
corresponds to the Brown-Henneaux boundary conditions so that excitations
are described by chiral movers. In the case of $k=1$, the boundary
gravitons fulfill the KdV equation and the asymptotic symmetry algebra
turns out to be infinite-dimensional, abelian and devoid of central
extensions. The latter feature also holds for the remaining cases
that describe the hierarchy ($k>1$). Our boundary conditions then
provide a gravitational dual of two noninteracting left and right
KdV movers, and hence, boundary gravitons possess anisotropic Lifshitz
scaling with dynamical exponent $z=2k+1$. Remarkably, despite spacetimes
solving the field equations are locally AdS, they possess anisotropic
scaling being induced by the choice of boundary conditions. As an
application, the entropy of a rotating BTZ black hole is precisely
recovered from a suitable generalization of the Cardy formula that
is compatible with the anisotropic scaling of the chiral KdV movers
at the boundary, in which the energy of AdS spacetime with our boundary
conditions depends on $z$ and plays the role of the central charge.
The extension of our boundary conditions to the case of higher spin gravity and its link with different classes of integrable systems is also briefly addressed.}

\makeatother

\begin{document}
\maketitle \flushbottom

\section{Introduction}

In the study of the asymptotic structure of spacetime it is customary
and natural to assume that the values of the lapse and shift functions,
describing the deformations of spacelike surfaces, are held fixed
to be constant at infinity \cite{Regge-Teitelboim,Henneaux:1985tv,Brown:1986nw,Henneaux:1985ey}.
Indeed, this choice ensures that observables, as the conserved charges,
are measured with respect to fixed time and length scales. Although,
this is certainly a reasonable and useful practice, it is not strictly
a necessary one. Here we explore some of the consequences of choosing
the time and length scales at the boundary in a non standard way,
so that the corresponding Lagrange multipliers are fixed at infinity
by a precise functional dependence on the dynamical fields. Afterwards,
we focus in the case of General Relativity with negative cosmological
constant in three spacetime dimensions. The standard analysis in this
case was performed in the metric formalism by Brown and Henneaux \cite{Brown:1986nw}.
In order to extend these results it turns out to be simpler to work
in terms of two independent $sl\left(2,\mathbb{R}\right)$ gauge fields,
$A^{\pm}=\omega\pm\frac{e}{\ell}$, where $\omega$ and $e$ stand
for the spin connection and the dreibein, respectively, so that General
Relativity can be formulated as a Chern-Simons theory \cite{AT,W}.
Following the lines of \cite{CHvD} the asymptotic fall--off of the
gauge fields can be written as 
\begin{equation}
A^{\pm}=g_{\pm}^{-1}\left(d+a^{\pm}\right)g_{\pm}\ ,\label{eq:Amn}
\end{equation}
so that the group elements $g_{\pm}=e^{\pm\log\left(r/\ell\right)L_{0}}$
entirely capture the radial dependence, and the components of the
auxiliary connections $a^{\pm}=a_{\varphi}^{\pm}d\varphi+a_{t}^{\pm}dt$,
depend only on time and the angular coordinate. According to the analysis
in \cite{Henneaux:2013dra,Bunster:2014mua}, the asymptotic behaviour
of the gauge fields is generically determined by 
\begin{eqnarray}
a_{\varphi}^{\pm} & = & L_{\pm1}-\frac{1}{4}\mathcal{L}_{\pm}L_{\mp1}\;\;;\;\;a_{t}^{\pm}=\pm\Lambda^{\pm}\left[\mu^{\pm}\right]\;,\label{eq:aphimn-atmn}
\end{eqnarray}
where 
\begin{equation}
\Lambda^{\pm}\left[\mu^{\pm}\right]=\mu^{\pm}\left(L_{\pm1}-\frac{1}{4}\mathcal{L}_{\pm}L_{\mp1}\right)\mp\mu^{\pm\prime}L_{0}+\frac{1}{2}\mu^{\pm\prime\prime}L_{\mp1}\;,\label{eq:Lambda}
\end{equation}
and $\mathcal{L}^{\pm}$, $\mu^{\pm}$ stand for arbitrary functions
of $t$, $\varphi$. \footnote{Here $L_{i}$ span each copy of $sl\left(2,\mathbb{R}\right)$, while
 $\ell$, $G$ denote the AdS radius and the Newton constant, respectively,
so that $\kappa=\frac{\ell}{4G}$. Hereafter, dot and prime stand
for derivatives with respect to $t$ and $\varphi$.}

The asymptotic form of the field equations $F^{\pm}=dA^{\pm}+A^{\pm}\land A^{\pm}=0$,
then reduces to 
\begin{equation}
\dot{\mathcal{L}}_{\pm}:=\pm{\cal D}^{\pm}\mu^{\pm}\;,\label{eq:FEmn}
\end{equation}
with 
\begin{equation}
{\cal D}^{\pm}:=\left(\partial_{\varphi}\mathcal{L}_{\pm}\right)+2\mathcal{L}_{\pm}\partial_{\varphi}-2\partial_{\varphi}^{3}\;.
\end{equation}

It is thus clear that in the reduced phase space, $\mathcal{L}_{\pm}$
describe the dynamical fields while $\mu^{\pm}$ are Lagrange multipliers.

It must be emphasized here that the set of boundary conditions is
not yet specified at this step, because in order to do that, one needs
to provide the precise form in which the Lagrange multipliers $\mu^{\pm}$
are fixed at infinity.

In the standard approach \cite{CHvD} the Lagrange multipliers are
chosen to be fixed at the boundary according to $\mu^{\pm}=1$, in
agreement with Brown and Henneaux \cite{Brown:1986nw}. These boundary
conditions can also be slightly generalized as in \cite{Henneaux:2013dra,Bunster:2014mua}
so that the Lagrange multipliers are chosen as $\mu^{\pm}=\mu_{0}^{\pm}\left(t,\varphi\right)$,
where $\mu_{0}^{\pm}$ stand for arbitrary functions of $t$, $\varphi$
that are held fixed at the boundary without variation, i.e., $\delta\mu_{0}^{\pm}=0$.

In the next section we explore the set of different possible choices
of Lagrange multipliers $\mu^{\pm}$ that are allowed by consistency
of the action principle.

\section{Specifying generic boundary conditions\label{sec:Specifying-generic-boundary}}

The action principle for General Relativity in terms of $sl\left(2,\mathbb{R}\right)$
gauge fields acquires the form
\begin{equation}
I=I_{CS}\left[A^{+}\right]-I_{CS}\left[A^{-}\right]\;,
\end{equation}
where $I_{CS}\left[A^{\pm}\right]$ stands for the Chern-Simons action.
For the remaining analysis it is useful to split the connection along
the spacelike and timelike components, $A^{\pm}=A_{i}^{\pm}dx^{i}+A_{t}^{\pm}dt$,
so that 
\begin{equation}
I_{CS}\left[A^{\pm}\right]=-\frac{\kappa}{4\pi}\int dtd^{2}x\varepsilon^{ij}\left\langle A_{i}^{\text{\ensuremath{\pm}}}\dot{A}_{j}^{\text{\ensuremath{\pm}}}-A_{t}^{\text{\ensuremath{\pm}}}F_{ij}^{\text{\ensuremath{\pm}}}\right\rangle +B_{\infty}^{\pm}\ .
\end{equation}
Here the bracket corresponds to the trace in the fundamental representation
of $sl\left(2,\mathbb{R}\right)$, and $B_{\infty}^{\pm}$ stand for
suitable boundary terms that are needed in order to ensure that the
action principle attains an extremum everywhere. Taking the variation
of the action with respect to the gauge fields, one finds that it
vanishes provided the curvatures $F^{\pm}$ also do in the bulk, while
for the asymptotic fall--off described by \eqref{eq:Amn}, \eqref{eq:aphimn-atmn},
the variation of the boundary terms is found to be given by
\[
\delta B_{\infty}^{\pm}=\mp\frac{\kappa}{8\pi}\int dtd\varphi\mu^{\pm}\delta\mathcal{L}_{\pm}\;.
\]
Therefore, since $sl\left(2,\mathbb{R}\right)$ gauge fields are assumed
to be independent, the action principle attains a bona fide extremum
provided the following integrability conditions are fulfilled: 
\begin{equation}
\delta^{2}B_{\infty}^{\pm}=\mp\frac{\kappa}{8\pi}\int dtd\varphi\delta\mu^{\pm}\wedge\delta\mathcal{L}_{\pm}=0\;.
\end{equation}
The integrability conditions are then solved by
\begin{equation}
\mu^{\pm}=\frac{\delta H^{\pm}}{\delta{\cal L}_{\pm}}\;,\label{eq:MudeltaHdeltaL}
\end{equation}
where $H^{\pm}$ can be assumed to correspond to arbitrary functionals
of ${\cal L}_{\pm}$ and their derivatives, i.e., $H^{\pm}=\int d\phi{\cal H}^{\pm}\left[{\cal L}_{\pm},{\cal L}_{\pm}^{\prime},{\cal L}_{\pm}^{\prime\prime},\cdots\right]$,
and hence, the boundary terms integrate as
\[
B_{\infty}^{\pm}=\mp\frac{\kappa}{8\pi}\int dtd\varphi{\cal H}^{\pm}\;.
\]

One then concludes that the boundary conditions become completely
determined once the functionals $H^{\pm}$ are specified at the boundary.
Consequently, the asymptotic form of the Lagrange multipliers $\mu^{\pm}$
is determined by eq. \eqref{eq:MudeltaHdeltaL}, which guarantees
the integrability of the boundary term required by consistency of
the action principle.

\subsection{Asymptotic symmetries and conserved charges}

Once the boundary conditions are generically specified through the
choice of $H^{\pm}$, one already possesses all what is needed in
order to study the asymptotic structure. By virtue of \eqref{eq:Amn}
the analysis of the asymptotic symmetries can be directly performed
in terms of the auxiliary connections $a^{\pm}$. We then look for
the subset of gauge transformations $\delta a^{\pm}=d\eta^{\pm}+\left[a^{\pm},\eta^{\pm}\right]$
that preserve their form, given by \eqref{eq:aphimn-atmn}. The asymptotic
form of $a_{\varphi}^{\pm}$ is maintained for gauge transformations
spanned by $\eta^{\pm}=\Lambda^{\pm}\left[\varepsilon^{\pm}\right]$,
where $\varepsilon^{\pm}=\varepsilon^{\pm}\left(t,\varphi\right)$,
and $\Lambda^{\pm}$ is defined in \eqref{eq:Lambda}, provided the
transformation law of the dynamical fields is given by
\begin{equation}
\delta\mathcal{L}_{\pm}={\cal D}^{\pm}\varepsilon^{\pm}.\label{eq:deltaLmn}
\end{equation}
Preserving the asymptotic form of $a_{t}^{\pm}$ then implies the
following conditions: 
\begin{equation}
\delta\mu^{\pm}=\pm\dot{\varepsilon}^{\pm}+\varepsilon^{\pm}\mu^{\pm\prime}-\mu^{\pm}\varepsilon^{\pm\prime}\;.\label{eq:deltaMu}
\end{equation}
Therefore, since the form of the Lagrange multipliers $\mu^{\pm}$
is determined by $H^{\pm}$ according to \eqref{eq:MudeltaHdeltaL},
the latter condition implies that the time derivatives of the parameters
$\varepsilon^{\pm}$ associated to the asymptotic symmetries have
to fulfill 
\begin{equation}
\dot{\varepsilon}^{\pm}=\pm\frac{\delta}{\delta\mathcal{L}_{\pm}}\int d\phi\frac{\delta H^{\pm}}{\delta\mathcal{L}_{\pm}}{\cal D}^{\pm}\varepsilon^{\pm}\;,\label{eq:ecsilonpunto}
\end{equation}
which implies that $\varepsilon^{\pm}$ generically acquire a nontrivial
dependence on the dynamical fields ${\cal L}_{\pm}$ and their derivatives.

Besides, in the canonical approach \cite{Regge-Teitelboim}, the variation
of the generators of the asymptotic symmetries is readily found to
be given by 
\begin{equation}
\delta Q^{\pm}\left[\varepsilon^{\pm}\right]=-\frac{\kappa}{8\pi}\int d\varphi\varepsilon^{\pm}\delta{\cal L}_{\pm}\;.\label{eq:can}
\end{equation}
It is then worth pointing out that eq. \eqref{eq:ecsilonpunto} guarantees
that the variation of the canonical generators is conserved in time
($\delta\dot{Q}^{\pm}=0$) on--shell.

However, in order to integrate the variation of the canonical generators
of the asymptotic symmetries in \eqref{eq:can}, one needs to know
the general solution of eq. \eqref{eq:ecsilonpunto}, which for a
generic choice of boundary conditions specified by $H^{\pm}$, turns
out to be a very hard task.

Nonetheless, this can always be done for the particular cases of asymptotic
Killing vectors $\partial_{\varphi}$ or $\partial_{t}$, when they
belong to the asymptotic symmetries. Indeed, in that cases, the angular
momentum reads
\begin{equation}
J=Q\left[\partial_{\varphi}\right]=\frac{\kappa}{8\pi}\int d\varphi\left({\cal L}_{+}-{\cal L}_{-}\right)\;,
\end{equation}
while the variation of the total energy, given by
\begin{equation}
\delta E=\delta Q\left[\partial_{t}\right]=\frac{\kappa}{8\pi}\int d\varphi\left(\mu^{+}\delta{\cal L}_{+}+\mu^{-}\delta{\cal L}_{-}\right)\;,
\end{equation}
by virtue of \eqref{eq:MudeltaHdeltaL}, integrates as
\begin{equation}
E=\frac{\kappa}{8\pi}\left(H^{+}+H^{-}\right)\;.\label{eq:E}
\end{equation}

In order to carry out the complete analysis of the asymptotic structure,
concrete choices of boundary conditions have then to be given.

\section{Selected choices of boundary conditions\label{sec:Selected-concrete-choices}}

A sensible criterium to fix the explicit form of $H^{\pm}$ turns
out to allow as much asymptotic symmetries as possible, which amounts
to know the general solution of \eqref{eq:ecsilonpunto} for arbitrary
values of the dynamical fields and their derivatives. Indeed, an infinite
number of asymptotic symmetries is certainly welcome because it helps
in order to explicitly find the space of solutions that fulfill the
boundary conditions. These criteria are certainly met in the cases
that $H^{\pm}$ define integrable systems. In what follows, we provide
a few (but still infinite) number of examples with the desired features.

\subsection{$k=0$: chiral movers (Brown--Henneaux)}

One of the simplest possible choices of boundary conditions corresponds
to the ones of Brown and Henneaux \cite{Brown:1986nw}. As aforementioned,
these boundary conditions are specified by choosing $\mu_{\left(0\right)}^{\pm}=1$,
which according to \eqref{eq:MudeltaHdeltaL}, amounts to set $H_{\left(0\right)}^{\pm}=\int d\varphi{\cal H}_{\left(0\right)}^{\pm}$,
with
\begin{equation}
{\cal H}_{\left(0\right)}^{\pm}:={\cal L}_{\pm}\;.\label{eq:Hcal_0}
\end{equation}
In this case the field equations \eqref{eq:FEmn} reduce to
\begin{equation}
\dot{{\cal L}}_{\pm}=\pm{\cal L}_{\pm}^{\prime}\;,
\end{equation}
describing chiral movers. Analogously, eq. \eqref{eq:ecsilonpunto}
reads 
\begin{equation}
\dot{{\cal \varepsilon}}^{\pm}=\pm\varepsilon^{\pm\prime}\;,
\end{equation}
so that the parameters that describe the asymptotic symmetries do
not depend on the dynamical fields. Therefore, the variation of the
canonical generators in eq. \eqref{eq:can} directly integrates as
\begin{equation}
Q^{\pm}\left[\varepsilon^{\pm}\right]=-\frac{\kappa}{8\pi}\int d\varphi\varepsilon^{\pm}{\cal L}_{\pm}\;.
\end{equation}
The algebra of the global charges can then be readily obtained from
$\left\{ Q\left[\varepsilon_{1}\right],Q\left[\varepsilon_{2}\right]\right\} =\delta_{\varepsilon_{2}}Q\left[\varepsilon_{1}\right]$,
which by virtue of the transformation law of the dynamical fields
in \eqref{eq:deltaLmn} reduces to two independent copies of the Virasoro
algebra with the Brown--Henneaux central extension.

\subsection{$k=1$: KdV movers}

A different simple choice of boundary conditions is given by $\mu_{\left(1\right)}^{\pm}={\cal L}_{\pm}$,
which corresponds to setting $H_{\left(1\right)}^{\pm}=\int d\varphi{\cal H}_{\left(1\right)}^{\pm}$
\begin{equation}
{\cal H}_{\left(1\right)}^{\pm}:=\frac{1}{2}{\cal L}_{\pm}^{2}\;.\label{eq:Hcal_1}
\end{equation}
For this case the field equations \eqref{eq:FEmn} imply that left
and right movers are described by the KdV equation, i.e.,
\begin{equation}
\dot{\mathcal{L}}_{\pm}=\pm\left(3\mathcal{L}_{\pm}{\cal L}_{\pm}^{\prime}-2{\cal L}_{\pm}^{\prime\prime\prime}\right)\;.\label{eq:FEk1}
\end{equation}
The parameters associated to the asymptotic symmetries are subject
to fulfill eq. \eqref{eq:ecsilonpunto}, which here reduces to
\begin{equation}
\dot{\varepsilon}^{\pm}=\pm\left(3\mathcal{L}_{\pm}\partial_{\varphi}\varepsilon^{\pm}-2\partial_{\varphi}^{3}\varepsilon^{\pm}\right)\;,\label{eq:epsilondt k1}
\end{equation}
and since the KdV equation corresponds to an integrable system, we
know its general solution assuming that $\varepsilon^{\pm}$ are local
functions of $\mathcal{L}_{\pm}$ and their spatial derivatives. It
is given by a linear combination of the form
\begin{equation}
\varepsilon^{\pm}=\sum_{j=0}^{\infty}\eta_{\left(j\right)}^{\pm}R_{\left(j\right)}^{\pm}\;,\label{eq:EpsilonKdV}
\end{equation}
where $\eta_{\left(j\right)}^{\pm}$ are constants and $R_{\left(j\right)}^{\pm}$
stand for the Gelfand--Dikii polynomials. They can be defined through
the following recursion relation\footnote{For later convenience, we have chosen the factor in \eqref{eq:RR}
such that the polynomials become normalized according to $R_{\left(j\right)}={\cal L}^{j}+\cdots$,
where the ellipsis refers to terms that depend on derivatives of ${\cal L}$.}:
\begin{equation}
\partial_{\varphi}R_{\left(j+1\right)}^{\pm}=\frac{j+1}{2j+1}{\cal D}^{\pm}R_{\left(j\right)}^{\pm}\;,\label{eq:RR}
\end{equation}
and they fulfill 
\begin{equation}
R_{\left(j\right)}^{\pm}=\frac{\delta H_{\left(j\right)}^{\pm}}{\delta{\cal L}_{\pm}}\;.
\end{equation}
In particular, according to eqs. \eqref{eq:Hcal_0} and \eqref{eq:Hcal_1},
$R_{\left(0\right)}^{\pm}=\mu_{\left(0\right)}^{\pm}=1$, and $R_{\left(1\right)}^{\pm}=\mu_{\left(1\right)}^{\pm}={\cal L}_{\pm}$. 

Therefore, for an arbitrary asymptotic symmetry, being spanned by
\eqref{eq:EpsilonKdV}, the variation of the canonical generators
integrates as
\begin{equation}
Q^{\pm}\left[\varepsilon^{\pm}\right]=-\frac{\kappa}{8\pi}\sum_{j=0}^{\infty}\eta_{\left(j\right)}^{\pm}H_{\left(j\right)}^{\pm}\;.\label{eq:QepsilonHk}
\end{equation}
The algebra of the canonical generators is then found to be an abelian
one and devoid of central extensions, which goes by hand with the
well-known fact that the conserved charges of an integrable system,
as it is the case of KdV, are in involution.

In particular, the first four conserved charges of the series are
explicitly given by
\begin{align}
H_{\left(0\right)}^{\pm} & =\int d\varphi{\cal L}_{\pm}\;\;,\;\;H_{\left(1\right)}^{\pm}=\int d\varphi\frac{1}{2}{\cal L}_{\pm}^{2}\;\;,\;\;H_{\left(2\right)}^{\pm}=\int d\varphi\frac{1}{3}\left({\cal L}_{\pm}^{3}+2{\cal L}_{\pm}^{\prime2}\right)\;,\nonumber \\
H_{\left(3\right)}^{\pm} & =\int d\varphi\frac{1}{4}\left(\mathcal{L}_{\pm}^{4}+8\mathcal{L}_{\pm}\mathcal{L}_{\pm}^{\prime2}+\frac{16}{5}\mathcal{L}_{\pm}^{\prime\prime2}\right)\;,
\end{align}
and it is also worth pointing out that, for this choice of boundary
conditions, according to \eqref{eq:E}, the total energy of a gravitational
configuration is given by the sum of the energies of left and right
KdV movers, i.e., 
\begin{equation}
E=\frac{\kappa}{16\pi}\int d\varphi\left({\cal L}_{+}^{2}+{\cal L}_{-}^{2}\right)\;.
\end{equation}

\subsection{Generic $k$: KdV hierarchy\label{sub:Generic-:-KdV}}

One is then naturally led to extend the previous analysis through
a new family of boundary conditions that is labeled by a nonnegative
integer $k$, so that the Brown--Henneaux boundary conditions as well
as the ones that describes KdV movers are recovered for $k=0$,$1$,
respectively. The boundary conditions are proposed to be such that
the Lagrange multipliers are given by
\begin{equation}
\mu_{\left(k\right)}^{\pm}=R_{\left(k\right)}^{\pm}=\frac{\delta H_{\left(k\right)}^{\pm}}{\delta{\cal L}_{\pm}}\;,\label{eq:Mu-k}
\end{equation}
so that the field equations \eqref{eq:FEmn} now describe left and
right movers that evolve according to the $k$-th representative of
the KdV hierarchy: 
\begin{equation}
\dot{\mathcal{L}}_{\pm}=\pm{\cal D}^{\pm}R_{\left(k\right)}^{\pm}\;.\label{eq:FE k}
\end{equation}
The asymptotic symmetries are then spanned by parameters $\varepsilon^{\pm}$
that fulfill \eqref{eq:ecsilonpunto} with $H^{\pm}=H_{\left(k\right)}^{\pm}$.

Note that in the case of boundary conditions with $k>1$, the field
equations \eqref{eq:FE k} as well as the conditions on the parameters
in \eqref{eq:ecsilonpunto} become severely modified as compared with
the case of $k=1$ (see eqs. \eqref{eq:FEk1} and \eqref{eq:epsilondt k1}).
Nonetheless, the remarkable properties of the Gelfand--Dikii polynomials
imply that the general solution of \eqref{eq:ecsilonpunto} for $k>1$
is described precisely by the same series as in the case of $k=1$,
i.e., if $\varepsilon^{\pm}$ are assumed to be local functions of
$\mathcal{L}_{\pm}$ and their spatial derivatives, the parameters
are given by \eqref{eq:EpsilonKdV}. Consequently, the corresponding
canonical generators are precisely given by eq. \eqref{eq:QepsilonHk},
that possess an infinite dimensional abelian algebra with no central
extensions. However, for the choice of boundary conditions described
here, a gravitational configuration possesses a total energy that
corresponds to the sum of the energies of left and right movers that
evolve according to the $k$-th representative of the KdV hierarchy,
given by $E=E_{+}+E_{-}$, with
\begin{equation}
E_{\pm}=\frac{\kappa}{8\pi}H_{\left(k\right)}^{\pm}\;.\label{eq:Delta-k}
\end{equation}

An interesting remark is in order. For a generic choice of the integer
that labels the boundary conditions, given by $k$, the ``boundary
gravitons'' that fulfill the field equations \eqref{eq:FE k} possess
an anisotropic Lifshitz scaling that is characterized by a dynamical
exponent 
\begin{equation}
z=2k+1\;.\label{eq:dynamical exponent}
\end{equation}
This is because our boundary conditions make the field equations \eqref{eq:FE k}
to be invariant under\footnote{In order to explicitly check the invariance of the field equations
under the anisotropic scaling, it is useful to take into account that
under \eqref{eq:Lif-scaling}, the Gelfand--Dikii polynomials scale
as $R_{\left(k\right)}\rightarrow\lambda^{-2k}R_{\left(k\right)}$.} 
\begin{equation}
t\rightarrow\lambda^{z}t\;\;,\;\;\varphi\rightarrow\lambda\varphi\;\;,\;\;{\cal L}_{\pm}\rightarrow\lambda^{-2}{\cal L}_{\pm}\;.\label{eq:Lif-scaling}
\end{equation}
It is then worth highlighting that, although spacetimes that solve
the field equations are locally AdS, they remarkably inherit an anisotropic
scaling that is induced by our choice of boundary conditions in \eqref{eq:Mu-k}.
Indeed, the corresponding line elements are manifestly invariant under
\eqref{eq:Lif-scaling}, provided the radial coordinate scales as
$r\rightarrow\lambda^{-1}r$. This can be explicitly seen in section
\ref{sec:Summary-of-results}.

As it is explained in the next section, the anisotropic Lifshitz scaling
yields interesting consequences concerning the asymptotic growth of
the number of states in the context of black hole entropy.

\section{BTZ black hole with selected boundary conditions: global charges
and thermodynamics}

The BTZ black hole \cite{BTZ,BHTZ} fits within the choice of boundary
conditions in \eqref{eq:Mu-k} for an arbitrary nonnegative integer
$k$. Indeed, this class of configurations is described by constant
${\cal L}_{\pm}$, which trivially solves the field equations \eqref{eq:FE k}
that correspond to the $k$-th representative of KdV hierarchy. It
is worth noting that once the spacetime metric is reconstructed from
\eqref{eq:Amn} and expressed in an ADM decomposition, it acquires
a similar form as in the standard case, but where the lapse and shift
are now described in terms of $\mu_{\left(k\right)}^{\pm}={\cal L}_{\pm}^{k}$
(see section \ref{sec:Summary-of-results}). AdS spacetime is then
recovered for ${\cal L}_{\pm}=-1$.

Consequently, the energy associated to left and right movers in the
case of the BTZ black hole can be directly obtained from \eqref{eq:Delta-k},
which in terms of the dynamical exponent \eqref{eq:dynamical exponent},
reads
\begin{equation}
E_{\pm}=\frac{\kappa}{2}\frac{1}{z+1}\mathcal{L}_{\pm}^{\frac{z+1}{2}}\;.\label{eq:EpmBTZ}
\end{equation}
Note that for our boundary conditions the corresponding left and right
energies of AdS spacetime manifestly depend on the dynamical exponent,
so that they are given by
\begin{equation}
E_{0}^{\pm}\left[z\right]=\frac{\kappa}{2}\frac{1}{z+1}\left(-1\right)^{\frac{z+1}{2}}\;,\label{eq:EpmAdS}
\end{equation}
and turn out to be positive or negative in the case of even or odd
values of $k$, respectively.

It is also worth highlighting that the Bekenstein-Hawking entropy
\begin{equation}
S=\frac{A}{4G}=\pi\kappa\left(\sqrt{\mathcal{L}_{+}}+\sqrt{\mathcal{L}_{-}}\right)\;,
\end{equation}
once expressed in terms of the extensive variables, given by the left
and right energies in \eqref{eq:EpmBTZ}, reads
\begin{equation}
S=\pi\kappa\left(\frac{2}{\kappa}\left(z+1\right)\right)^{\frac{1}{z+1}}\left(E_{+}^{\frac{1}{z+1}}+E_{-}^{\frac{1}{z+1}}\right)\;.\label{eq:S_Em,En}
\end{equation}
 \bigskip{}
In terms of left and right temperatures $T_{\pm}=\beta_{\pm}^{-1}$,
given by\footnote{Left and right temperatures are related to the Hawking temperature
according to $\frac{1}{T}=\frac{1}{2}\left(\frac{1}{T_{+}}+\frac{1}{T_{-}}\right)$.} 
\begin{equation}
\beta_{\pm}=\frac{\partial S}{\partial E_{\pm}}=2\pi\left(\frac{2}{\kappa}\left(z+1\right)E_{\pm}\right)^{-\frac{z}{z+1}}\;,\label{eq:BetamnEmn}
\end{equation}
the black hole entropy acquires the form
\begin{equation}
S=\frac{\kappa}{2}\left(2\pi\right)^{1+\frac{1}{z}}\left(T_{+}^{\frac{1}{z}}+T_{-}^{\frac{1}{z}}\right)\;.\label{eq:EntropyT-z}
\end{equation}
Remarkably, the black hole entropy not only acquires the expected
dependence on the energy or the temperature of noninteracting left
and right movers of a field theory with Lifshitz scaling in two dimensions,
see e.g., \cite{MarikaTaylor-non-relativistic,Bertoldi:2009vn,Bertoldi:2009dt,DHoker-Kraus,Hartnoll,Marika-Taylor-Lifshitz-holo},
but it can actually be precisely recovered from a suitable generalization
of the Cardy formula in the case of anisotropic scaling, along the
lines of \cite{HernanTT}.

In the next section, we show how the results in \cite{HernanTT} extend
to the case of left and right movers with the same Lifshitz scaling,
whose corresponding ground state energies are allowed to depend on
the dynamical exponent $z$.

\section{Asymptotic growth of the number of states from anisotropic modular
invariance}

As explained in \cite{HernanTT}, thermal field theories with Lifshitz
scaling in two dimensions, defined on a torus parametrized by $0\leq\varphi<2\pi$,
$0\leq t_{E}<\beta$, where $t_{E}$ is the Euclidean time, naturally
possess a duality between low and high temperatures, given by
\begin{equation}
\frac{\beta}{2\pi}\rightarrow\left(\frac{2\pi}{\beta}\right)^{\frac{1}{z}}\;,
\end{equation}
 so that the partition function can be assumed to be invariant under
\begin{equation}
Z\left[\beta;z\right]=Z\left[\frac{\left(2\pi\right)^{1+\frac{1}{z}}}{\beta^{\frac{1}{z}}};\frac{1}{z}\right]\;.
\end{equation}
In the case of independent noninteracting left and right movers with
the same dynamical exponent $z$, the theory is defined on a torus
with modular parameter 
\begin{equation}
\tau=i\frac{\beta}{2\pi}\;,
\end{equation}
where $\beta$ stands for the complexification of left and right temperatures
$\beta_{\pm}$. The high/low temperature duality relation then reads
\begin{equation}
\tau\rightarrow\frac{i^{1+\frac{1}{z}}}{\tau^{\frac{1}{z}}}\;,
\end{equation}
and therefore, one can assume that the partition function fulfills
\begin{equation}
Z\left[\tau;z\right]=Z\left[i^{1+\frac{1}{z}}\tau^{-\frac{1}{z}};z^{-1}\right]\;.\label{eq: Z=00005B=00005D ModularInv}
\end{equation}
This is the anisotropic version of the well known S-modular invariance
for chiral movers, which for $z=1$ reduces to standard one in conformal
field theory \cite{Cardy,DiFrabcescolibro}.

If one assumes that the spectrum of left and right movers possesses
a gap, the high/low temperature duality allows to obtain the precise
value of the asymptotic growth of the number of states for fixed left
and right energies $\Delta_{\pm}$. The existence of a gap ensures
that at low temperatures the partition function turns out to be dominated
by the ground state. It is also assumed that the ground state is not
degenerate, being such that its left and right energies are negative
and given by $-\Delta_{0}^{\pm}\left[z\right]$, which generically
depend on the dynamical exponent. Therefore, the partition function
at low temperature approximates as 
\begin{equation}
Z[\tau;z]\approx e^{-2\pi i\left(\tau\Delta_{0}\left[z\right]-\bar{\tau}\bar{\Delta}_{0}\left[z\right]\right)}\;,
\end{equation}
so that in the high temperature regime, by virtue of \eqref{eq: Z=00005B=00005D ModularInv},
the partition function reads
\begin{equation}
Z[\tau;z]\approx e^{2\pi\left(\left(-i\tau\right)^{-\frac{1}{z}}\Delta_{0}\left[z^{-1}\right]+\left(i\bar{\tau}\right)^{-\frac{1}{z}}\bar{\Delta}_{0}\left[z^{-1}\right]\right)}\;.\label{eq:Z-Modu-2}
\end{equation}
Hence, at fixed energies $\Delta_{\pm}\gg\Delta_{0}^{\pm}\left[z\right]$,
the asymptotic growth of the number of states can be directly obtained
by evaluating \eqref{eq:Z-Modu-2} in the saddle point approximation,
which is described by an entropy given by
\begin{equation}
S=2\pi\left(z+1\right)\left[\left(\frac{\Delta_{0}\left[z^{-1}\right]}{z}\right)^{z}\Delta\right]^{\frac{1}{z+1}}+2\pi\left(z+1\right)\left[\left(\frac{\bar{\Delta}_{0}\left[z^{-1}\right]}{z}\right)^{z}\bar{\Delta}\right]^{\frac{1}{z+1}}\;.\label{eq:CardyLif}
\end{equation}
Note that the Cardy formula is recovered in the case of $z=1$, where
the role of the central charges is expressed through the lowest eigenvalues
of the shifted Virasoro operators $L_{0}\rightarrow L_{0}-\frac{c}{24}$,
see e.g., \cite{Cardy,Carlip:1999cy,Correa-Troncoso-martinez1,Correa-Troncoso-Martinez2}.

In terms of the (Lorentzian) left and right energies, the entropy
reads 
\begin{equation}
S=2\pi\left(z+1\right)\left[\left(\frac{\left|\Delta_{0}^{+}\left[z^{-1}\right]\right|}{z}\right)^{z}\Delta_{+}\right]^{\frac{1}{z+1}}+2\pi\left(z+1\right)\left[\left(\frac{\left|\Delta_{0}^{-}\left[z^{-1}\right]\right|}{z}\right)^{z}\Delta_{-}\right]^{\frac{1}{z+1}}\;,\label{eq:CardyLif-Lorentz}
\end{equation}
and from the first law in the canonical ensemble, $dS=\beta_{+}d\Delta_{+}+\beta_{-}d\Delta_{-}$,
one finds that left and right movers follow an anisotropic version
of the Stefan-Boltzmann law, given by
\begin{equation}
\Delta_{\pm}=\frac{1}{z}\left(2\pi\right)^{1+\frac{1}{z}}\left|\Delta_{0}^{\pm}\left[z^{-1}\right]\right|T_{\pm}^{1+\frac{1}{z}}\;,\label{eq:Anisotropic-StenfaH}
\end{equation}
which reduces to the standard one for $z=1$.

In terms of left and right temperatures the entropy \eqref{eq:CardyLif-Lorentz}
then reads
\begin{equation}
S=\left(2\pi\right)^{1+\frac{1}{z}}\left(1+\frac{1}{z}\right)\left(\left|\Delta_{0}^{+}\left[z^{-1}\right]\right|T_{+}^{\frac{1}{z}}+\left|\Delta_{0}^{-}\left[z^{-1}\right]\right|T_{-}^{\frac{1}{z}}\right)\;.\label{eq:Cardy-Lif-Lorentz-Tmn}
\end{equation}

It is then very remarkable that the Bekenstein--Hawking entropy of
the BTZ black hole, once expressed in terms of the energies associated
to left and right movers that evolve according to the field equations
of the $k$-th KdV hierarchy, given by \eqref{eq:S_Em,En}, is precisely
reproduced from \eqref{eq:CardyLif-Lorentz}. Indeed, this is the
case if one identifies the left and right energies of the field theory
with the ones of the black hole, i.e., $\Delta_{\pm}=E_{\pm}$, provided
the ground state energies correspond to the ones of AdS spacetime
with our boundary conditions, $\Delta_{0}^{\pm}\left[z\right]=-E_{0}^{\pm}\left[z\right]$,
where $E_{0}^{\pm}\left[z\right]$ is given by \eqref{eq:EpmAdS}.

Analogously, the anisotropic Stefan-Boltzmann law \eqref{eq:Anisotropic-StenfaH}
agrees with \eqref{eq:BetamnEmn}, as well as eq. \eqref{eq:Cardy-Lif-Lorentz-Tmn}
does with \eqref{eq:EntropyT-z}.

\section{Summary of results in terms of the spacetime metric\label{sec:Summary-of-results}}

For a generic choice of boundary conditions, specified by $\mu^{\pm}$
in \eqref{eq:MudeltaHdeltaL}, the asymptotic structure of the spacetime
metric can\emph{ }be reconstructed from the asymptotic form of the
$sl\left(2,\mathbb{R}\right)$ gauge fields given by \eqref{eq:Amn},
\eqref{eq:aphimn-atmn}. The fall--off of the metric in the asymptotic
region, $r\gg\ell$, reads
\begin{align}
g_{tt} & =-\left({\cal N}{}^{2}-\ell^{2}{\cal N}^{\varphi}{}^{2}\right)\frac{r^{2}}{\ell^{2}}+f_{tt}+{\cal O}\left(r^{-1}\right)\;,\nonumber \\
g_{tr} & =-{\cal N}^{\varphi}{}^{\prime}\frac{\ell^{2}}{r}+{\cal O}\left(r^{-4}\right)\;,\nonumber \\
g_{t\varphi} & ={\cal N}^{\varphi}r^{2}+f_{t\varphi}+{\cal O}\left(r^{-1}\right)\;,\label{eq:AsympMetric}\\
g_{rr} & =\frac{\ell^{2}}{r^{2}}+O\left(r^{-5}\right)\;,\nonumber \\
g_{\varphi\varphi} & =r^{2}+f_{\varphi\varphi}+{\cal O}\left(r^{-1}\right)\;,\nonumber \\
g_{r\varphi} & =O\left(r^{-3}\right)\;,\nonumber 
\end{align}
with 
\begin{equation}
\mu^{\pm}={\cal N}\ell^{-1}\pm{\cal N}^{\varphi}\;.\label{eq:NMumn}
\end{equation}
Therefore, in an ADM decomposition, the lapse and the shift asymptotically
behave as
\begin{align}
N^{\perp}= & \frac{r}{\ell}{\cal N}+{\cal O}\left(r^{-1}\right)\;,\\
N^{r}= & -r{\cal N}^{\varphi\prime}+{\cal O}\left(r^{-1}\right)\;,\\
N^{\varphi}= & {\cal N}^{\varphi}+{\cal O}\left(r^{-2}\right)\;,
\end{align}
and consequently, by virtue of \eqref{eq:NMumn}, they become determined
at the boundary according to eq. \eqref{eq:MudeltaHdeltaL}.

The functions $f_{\varphi\varphi}$, $f_{t\varphi}$, and $f_{tt}$
are given by 
\begin{align*}
f_{\varphi\varphi} & =\frac{\ell^{2}}{4}\left({\cal L}_{+}+{\cal L}_{-}\right)\;,\\
f_{t\varphi} & =-\frac{\ell^{2}}{2}{\cal N}^{\varphi\prime\prime}+f_{\varphi\varphi}{\cal N}^{\varphi}+\frac{\ell}{4}\left({\cal L}_{+}-{\cal L}_{-}\right){\cal N}\;,\\
f_{tt} & =\left(\frac{1}{\ell^{2}}{\cal N}{}^{2}-{\cal N}^{\varphi}{}^{2}\right)f_{\varphi\varphi}+2f_{t\varphi}{\cal N}^{\varphi}+\ell^{2}{\cal N}^{\varphi\prime2}-{\cal N}{\cal N}^{\prime\prime}\;.
\end{align*}

The asymptotic form of the metric \eqref{eq:AsympMetric} then implies
that the Einstein equations with negative cosmological constant in
vacuum are fulfilled provided
\begin{equation}
\dot{\mathcal{L}}_{\pm}=\pm{\cal D}^{\pm}\mu^{\pm}\;,\label{eq:Lpunto}
\end{equation}
in full agreement with \eqref{eq:FEmn}.

The asymptotic form of the metric \eqref{eq:AsympMetric} is mapped
into itself under asymptotic Killing vectors $\xi^{\mu}$ that fulfill
$\delta_{\xi}g_{\mu\nu}=\mathscr{L}_{\xi}g_{\mu\nu}$, whose components
are given by 
\begin{align}
\xi^{t} & =\frac{\ell}{2{\cal N}}\left[\varepsilon^{+}+\varepsilon^{-}+\frac{\ell^{2}}{2{\cal N}r^{2}}\left({\cal N}\left(\varepsilon^{+}+\varepsilon^{-}\right)^{\prime\prime}-{\cal N}{}^{\prime\prime}\left(\varepsilon^{+}+\varepsilon^{-}\right)\right)\right]+{\cal O}\left(r^{-4}\right)\;,\\
\xi^{r} & =-\frac{1}{2{\cal N}}\left[\left(\varepsilon^{+}-\varepsilon^{-}\right)^{\prime}{\cal N}-\ell{\cal N}^{\varphi}{}^{\prime}\left(\varepsilon^{+}+\varepsilon^{-}\right)\right]r\nonumber \\
 & +\frac{\ell^{3}{\cal N}^{\varphi\prime}}{4{\cal N}r}\left[\left(\varepsilon^{+}+\varepsilon^{-}\right)^{\prime\prime}-\left(\varepsilon^{+}+\varepsilon^{-}\right)\frac{{\cal N}^{\prime\prime}}{{\cal N}}\right]\frac{1}{r}+{\cal O}\left(r^{-2}\right)\;,\\
\xi^{\varphi} & =\frac{1}{2{\cal N}}\left[\left(\varepsilon^{+}-\varepsilon^{-}\right){\cal N}-\ell\left(\varepsilon^{+}+\varepsilon^{-}\right){\cal N}^{\varphi}\right]-\frac{\ell^{2}}{2{\cal N}r^{2}}\left[\left(\varepsilon^{+}+\varepsilon^{-}\right)^{\prime\prime}{\cal N}+\ell\left(\varepsilon^{+}-\varepsilon^{-}\right)^{\prime\prime}{\cal N}^{\varphi}\right.\nonumber \\
 & \left.-\frac{\ell}{{\cal N}}\left(\varepsilon^{+}+\varepsilon^{-}\right)\left({\cal N}{\cal N}^{\varphi\prime\prime}+{\cal N}^{\prime\prime}{\cal N}^{\varphi}\right)\right]+{\cal O}\left(r^{-4}\right)\;,
\end{align}
with $\varepsilon^{\pm}=\varepsilon^{\pm}\left(t,\varphi\right)$,
provided ${\cal L}_{\pm}$ and $\mu^{\pm}$ transform precisely according
to eqs. \eqref{eq:deltaLmn} and \eqref{eq:deltaMu}, respectively.
Note that consistency of \eqref{eq:deltaMu} implies that the time
derivative of $\varepsilon^{\pm}$ fulfills \eqref{eq:ecsilonpunto},
which means that these parameters in general depend on ${\cal L}_{\pm}$
and their derivatives.

The variation of the global charges associated to the asymptotic symmetries
can then be obtained in the canonical approach \cite{Regge-Teitelboim},
and they are found to agree with \eqref{eq:can}.

The general solution of the field equations that fulfills our boundary
conditions \eqref{eq:AsympMetric} is described by spacetime metrics
that in an ADM decomposition read, 
\begin{equation}
ds^{2}=-\left(N^{\perp}\right)^{2}dt^{2}+g_{ij}\left(N^{i}dt+dx^{i}\right)\left(N^{j}dt+dx^{j}\right)\;,\label{eq:general solution}
\end{equation}
where the spacelike geometry is given by
\begin{equation}
dl^{2}=g_{ij}dx^{i}dx^{j}=\frac{\ell^{2}}{r^{2}}\left[dr^{2}+\ell^{2}\left(\frac{r^{2}}{\ell^{2}}+\frac{1}{4}{\cal L}_{+}\right)\left(\frac{r^{2}}{\ell^{2}}+\frac{1}{4}{\cal L}_{-}\right)d\varphi^{2}\right]\;,
\end{equation}
with the following shift and lapse functions
\begin{align}
N^{r} & =-r{\cal N}^{\varphi\prime}\;,\;N^{\varphi}=\mathcal{N}^{\varphi}+\frac{\left(\frac{r^{2}}{\ell^{2}}\mathcal{N}+\frac{1}{4}\mathcal{N}^{\prime\prime}\right)\left(\mathcal{L}_{+}-\mathcal{L}_{-}\right)-2\left(\frac{r^{2}}{\ell^{2}}+\frac{1}{8}\left(\mathcal{L}_{+}+\mathcal{L}_{-}\right)\right)\mathcal{N}^{\varphi\prime\prime}}{4\ell\left(\frac{r^{2}}{\ell^{2}}+\frac{1}{4}\mathcal{L}_{+}\right)\left(\frac{r^{2}}{\ell^{2}}+\frac{1}{4}\mathcal{L}_{-}\right)}\;,\\
N^{\perp} & =\frac{\ell\left[\left(\frac{r^{4}}{\ell^{4}}-\frac{1}{16}\mathcal{L}_{+}\mathcal{L}_{-}\right)\mathcal{N}+\frac{1}{2}\left(\frac{r^{2}}{\ell^{2}}+\frac{1}{8}\left(\mathcal{L}_{+}+\mathcal{L}_{-}\right)\right)\mathcal{N}^{\prime\prime}-\frac{\ell}{16}\left(\mathcal{L}_{+}-\mathcal{L}_{-}\right)\mathcal{N}^{\varphi\prime\prime})\right]}{r\sqrt{\left(\frac{r^{2}}{\ell^{2}}+\frac{1}{4}\mathcal{L}_{+}\right)\left(\frac{r^{2}}{\ell^{2}}+\frac{1}{4}\mathcal{L}_{-}\right)}}\;,
\end{align}
respectively, where ${\cal L}_{\pm}$ satisfy eq. \eqref{eq:Lpunto}.

Therefore, for the choice of boundary conditions that is labelled
by a nonnegative integer $k$ in \eqref{eq:Mu-k}, the class of spacetimes
described by \eqref{eq:general solution} solves the Einstein equations
with negative cosmological constant in vacuum as long as the functions
$\mathcal{L}_{\pm}$ fulfill the field equations of left and right
movers for the $k$-th element of the KdV hierarchy \eqref{eq:Lpunto}. 

In this case, the line element \eqref{eq:general solution} turns
out to be manifestly invariant under the anisotropic (Lifshitz) scaling
transformation given by \eqref{eq:Lif-scaling}, provided the radial
coordinate scales as $r\rightarrow\lambda^{-1}r$.

In the particular case of constant $\mathcal{L}_{\pm}$, the metric
\eqref{eq:general solution} reduces to
\begin{align}
ds^{2} & =\ell^{2}\left[\frac{dr^{2}}{r^{2}}+\frac{\mathcal{L}_{+}}{4}\left(d\tilde{x}^{+}\right)^{2}+\frac{\mathcal{L}_{-}}{4}\left(d\tilde{x}^{-}\right)^{2}-\left(\frac{r^{2}}{\ell^{2}}+\frac{\ell^{2}\mathcal{L}_{+}\mathcal{L}_{-}}{16r^{2}}\right)d\tilde{x}^{+}d\tilde{x}^{-}\right]\;,
\end{align}
with 
\begin{equation}
d\tilde{x}^{\pm}=\mu^{\pm}dt\pm d\varphi\;,
\end{equation}
which corresponds to the BTZ black hole where the lapse and the shift
are determined by $\mu^{\pm}=\mathcal{L}_{\pm}^{k}$. The left and
right temperatures in \eqref{eq:BetamnEmn} can then be readily found
by requiring the Euclidean metric to be smooth at the horizon.

For a generic choice of boundary conditions, the lapse and the shift
can be obtained from $\mu^{\pm}=\frac{\delta H^{\pm}}{\delta\mathcal{L_{\pm}}}$,
and the Euclidean metric becomes regular for $\mu^{\pm}=\frac{2\pi}{{\cal \sqrt{L_{\pm}}}}$.

\section{Discussion}

We have shown that the dynamics of left and right movers that evolve
according to the field equations of $k$-th element of KdV hierarchy
can be fully geometrized. Indeed, the general solution of the three--dimensional
Einstein equations with negative cosmological constant with our boundary
conditions in \eqref{eq:Mu-k} is described by spacetime geometries
of the form \eqref{eq:general solution}, where $\mathcal{L}_{\pm}\left(t,\varphi\right)$
fulfill the field equations \eqref{eq:FE k}. Consequently, in this
framework, the parameters that characterize the $k$-th KdV equation
acquire a gravitational meaning.

Interestingly, different phenomena that have been observed in KdV
can then be interpreted in the context of gravitation and vice versa.

In particular, let us consider the simplest solutions of the $k$-th
KdV equations with $k\neq0$, being described by constants $\mathcal{L}_{\pm}$,
which possesses fixed values of left and right energies, determined
by $H_{\left(k\right)}^{\pm}$. One can then generate a generic solution
by acting on the simplest one with an arbitrary linear combination
of the remaining generators $H_{\left(j\right)}^{\pm}$ with $j\neq k$.
As it is well known in the case $k=1$, see e.g. \cite{Dauxois:2006zz},
the evolution in time of this generic configuration will settle down
to describe a superposition of left/right solitons (cnoidal waves)
with right/left dispersive waves, with the same energy of the original
configuration, since $\{H_{\left(k\right)}^{\pm},H_{\left(j\right)}^{\pm}\}=0$.

The gravitational interpretation of this phenomenon is the following:
the simplest configuration with constants $\mathcal{L}_{\pm}$ describes
a BTZ black hole. The generic solution is then obtained by acting
on the BTZ black hole with a generic element of the asymptotic symmetry
group globally. Since the asymptotic symmetry generators commute with
the Hamiltonian, this operation turns out to be a ``soft boost''
\cite{Daniel}. One then obtains an inequivalent gravitational configuration
that can be regarded as a black hole with ``soft gravitons'' on
it, in the sense of Hawking, Perry and Strominger \cite{Hawking}.
This is because the new configuration is generically endowed with
nontrivial global charges $H_{\left(j\right)}^{\pm}$, so that it
is not related with the original one by a pure gauge transformation.
Hence, since these global charges commute with the energy, they can
be properly cast as ``soft hair''. Nonetheless, one should check
whether the generic configuration could properly describe a black
hole, since a priori there is no guarantee that it possesses a regular
horizon in the standard sense.

\medskip{}

Besides, and quite remarkably, our choice of boundary conditions in
\eqref{eq:Mu-k} describes constant curvature spacetimes that are
locally AdS with anisotropic Lifshitz scaling with dynamical exponent
$z=2k+1$. This opens the possibility of studying nonrelativistic
holography along the lines of \cite{MarikaTaylor-non-relativistic,Hartnoll:2009sz,Hartnoll:2011fn,Marika-Taylor-Lifshitz-holo},
but without the need of bulk geometries described by asymptotically
Lifshitz spacetimes. In other words, Lifshitz scaling does not necessarily
requires the use of Lifshitz spacetimes. Indeed, the BTZ black hole
entropy with our boundary conditions, given by \eqref{eq:S_Em,En},
is successfully reproduced from the asymptotic growth of the number
of states of a field theory that describes left and right movers with
Lifshitz scaling with the same dynamical exponent $z$ in \eqref{eq:CardyLif}.
In the non--chiral case, the latter formula reduces to the one proposed
in \cite{HernanTT}, which also precisely reproduces the entropy of
different classes of asymptotically Lifshitz black holes \cite{HernanTT,Eloy,Hassaine1}\footnote{Different generalizations of the Cardy formula have also been found
for alternative scaling laws in three-dimensional spacetimes \cite{BarnichFlat,Bagchi-Detournay,Detournay-Hartman,shag,Hassaine3,Castro-Hoffman,Afshar-Detournay,Detournay-Celine}.}. A different link between asymptotically Lifshitz black holes and
a generalization of the KdV equation has been pointed out in \cite{Abdalla}.

\medskip{}

Note that in order to obtain the asymptotic growth of the number of
states in \eqref{eq:CardyLif}, neither the asymptotic symmetries
nor central charges were required. Indeed, the role of the central
charge in our case is played by the left and right energies $\Delta_{0}^{\pm}\left[z^{-1}\right]=-E_{0}^{\pm}\left[z^{-1}\right]$
of AdS spacetime, given by \eqref{eq:EpmAdS} with $z\rightarrow z^{-1}$.
In this context an interesting remark is worth to be mentioned. Note
that for our boundary conditions with odd values of $k$, $E_{0}^{\pm}\left[z\right]$
turns out to be positive. However, in these cases, when one performs
the Euclidean continuation, it is found that the Euclidean BTZ with
our boundary conditions is diffeomorphic to thermal AdS as in \cite{Carlip-Teitelboim,Maldacena-Strominger},
but with reversed orientation, which is a consequence of the fact
that the lapse of AdS spacetime reverses its sign. Therefore, left
and right energies of (Euclidean) thermal AdS possess an opposite
sign as compared with the Lorentzian ones $E_{0}^{\pm}\left[z\right]$
of AdS spacetime for odd $k$.\medskip{}

It would also be interesting to explore different possible choices
of boundary conditions for which the Lagrange multipliers depend on
the dynamical fields in different ways as compared with the ones considered
here. For instance, this is the case for the set of boundary conditions
that has been recently proposed in \cite{Daniel}, in which the Lagrange
multipliers $\mu^{\pm}$ depend non--locally on $\mathcal{L}_{\pm}$.
Indeed, up to the conventional normalization factor in \eqref{eq:RR},
one verifies that the recursion relation allows to recover the Brown-Henneaux
boundary conditions starting from the ones in \cite{Daniel}, because
the former stand for the kernel of $\partial_{\varphi}$, while the
latter corresponds to the kernel of the operator ${\cal D}$. Consequently,
the boundary conditions in \cite{Daniel} are associated with an anisotropic
scaling with $z=0$, that is consistent with labelling this set as
a member of an extended hierarchy with $k=-1/2$. Such kind of extensions
of the KdV hierarchy have been studied in e.g. \cite{FracKdV1,FracKdV2,FracKdV3,FracKdV4}.

A more conservative choice of boundary conditions corresponds to deform
the Hamiltonian of the $k$-th KdV hierarchy $H_{\left(k\right)}^{\pm}$
by a linear combination of the remaining generators of the asymptotic
symmetries, so that $H^{\pm}=H_{\left(k\right)}^{\pm}+\sum_{j\neq k}\xi_{\left(j\right)}^{\pm}H_{\left(j\right)}^{\pm},$
with $\xi_{\left(j\right)}^{\pm}$ constants. It is clear that in
this case the asymptotic symmetries remain the same, but nonetheless,
the Lifshitz scaling symmetry is lost. Note that in the context of
the AdS/CFT correspondence, this is interpreted as a multitrace deformation
of the dual theory \cite{Klebanov:1999tb,Witten:2001ua,Shomer}. The
particular case of $k=0$ then corresponds to a multitrace deformation
of the Brown-Henneaux boundary conditions, which has been recently
considered in \cite{de Boer}, in the context of generalized Gibbs
ensembles for which all of the additional charges $H_{\left(j\right)}^{\pm}$
possess nonvanishing chemical potentials $\xi_{\left(j\right)}^{\pm}$.

\medskip{}

Our analysis of generic choices of boundary conditions can be readily
generalized to the case of higher spin gravity in three-spacetime
dimensions. Some particular examples have already been reported in
\cite{Compere-Song,Gutperle1,Gutperle2}.

If one follows the lines explained in section \ref{sec:Specifying-generic-boundary},
one finds that the Lagrange multipliers are of the form $\mu_{i}=\mu_{i}\left(\mathcal{W}_{s}\right)$,
where $\mathcal{W}_{s}$, with $i,s=2,3,\cdots,N$, stand for the
spin-$s$ charges which appear in the component $a_{\varphi}$ of
the $sl\left(N,\mathbb{R}\right)$ gauge fields, while $\mu_{i}$
enter through $a_{t}$. Note that $a_{\varphi}$, $a_{t}$ form a
Lax pair. Hence, for instance, our proposal for the boundary conditions
in section \ref{sub:Generic-:-KdV}, in the case of $sl\left(3,\mathbb{R}\right)$
corresponds to choosing the Lagrange multipliers associated to the
spin-2 and spin-3 charges, given by $\mathcal{L}$ and $\mathcal{W}$,
respectively, according to $\mu_{{\cal L}}=R_{\left(k\right)}$ and
$\mu_{{\cal W}}=S_{\left(k\right)}$, where the doublet $\left(R_{\left(k\right)},S_{\left(k\right)}\right)$
stands for the generalized Gelfand-Dikii polynomials associated to
the Boussinesq hierarchy, see e.g., \cite{Drinfeld:1984qv,Das:1990td,Mathieu:1991et}.
The case of $k=0$ then reduces to the set of boundary conditions
proposed in \cite{Henneaux:2013dra,Bunster:2014mua}.

\acknowledgments We thank O. Fuentealba, H. González, M. Grigoriev,
D. Grumiller, C. Martínez, J. Matulich, W. Merbis, B. Oblak and M.
Pino for useful comments. We are indebted to C. Troessaert for enlightening
remarks and collaboration in an early stage of this work. This research
has been partially supported by Fondecyt grants Nº 11130260, 11130262,
1130658, 1161311. The Centro de Estudios Científicos (CECs) is funded
by the Chilean Government through the Centers of Excellence Base Financing
Program of Conicyt. 

\appendix
%dummy comment inserted by tex2lyx to ensure that this paragraph is not empty


\begin{thebibliography}{10}
\bibitem{Regge-Teitelboim}T.~Regge and C.~Teitelboim,   ``Role of Surface Integrals in the Hamiltonian Formulation of General Relativity,''   Annals Phys.\  {\bf 88}, 286 (1974).   doi:10.1016/0003-4916(74)90404-7   %%CITATION = doi:10.1016/0003-4916(74)90404-7;%%

\bibitem{Henneaux:1985tv}M.~Henneaux and C.~Teitelboim,   ``Asymptotically anti-De Sitter Spaces,''   Commun.\ Math.\ Phys.\  {\bf 98}, 391 (1985).   doi:10.1007/BF01205790   %%CITATION = doi:10.1007/BF01205790;%%

\bibitem{Brown:1986nw}J.~D.~Brown and M.~Henneaux, ``Central
Charges in the Canonical Realization of Asymptotic Symmetries: An
Example from Three-Dimensional Gravity,'' Commun.\ Math.\ Phys.\ \textbf{104},
207 (1986). %doi:10.1007/BF01211590   %%CITATION = doi:10.1007/BF01211590;%%

\bibitem{Henneaux:1985ey}M.~Henneaux,   ``Asymptotically Anti-de Sitter Universes In D = 3, 4 And Higher Dimensions,''
in Proceedings of the Fourth Marcel Grossmann Meeting on General Relativity, Rome 1985, R. Ruffini, ed., pp. 959-966. Elsevier Science Publishers B.V., 1986.
%%CITATION = INSPIRE-224274;%%

\bibitem{AT}A.~Achucarro and P.~K.~Townsend, ``A Chern-Simons
Action for Three-Dimensional anti-De Sitter Supergravity Theories,\textquotedblright \ Phys.\ Lett.\ B
\textbf{180}, 89 (1986). %%CITATION = PHLTA,B180,89;%%

\bibitem{W}E.~Witten, ``(2+1)-Dimensional Gravity as an Exactly
Soluble System, \textquotedblright \ Nucl.\ Phys.\ B \textbf{311},
46 (1988). %%CITATION = NUPHA,B311,46;%%

\bibitem{CHvD}O.~Coussaert, M.~Henneaux and P.~van Driel, ``The
Asymptotic dynamics of three-dimensional Einstein gravity with a negative
cosmological constant,'' Class.\ Quant.\ Grav.\ \textbf{12}, 2961
(1995) {[}gr-qc/9506019{]}. %%CITATION = GR-QC/9506019;%%

\bibitem{Henneaux:2013dra} M.~Henneaux, A.~Perez, D.~Tempo and
R.~Troncoso,``Chemical potentials in three-dimensional higher spin
anti-de Sitter gravity,'' JHEP \textbf{1312}, 048 (2013) {[}arXiv:1309.4362
{[}hep-th{]}{]}. %doi:10.1007/JHEP12(2013)048   %%CITATION = doi:10.1007/JHEP12(2013)048;%%

\bibitem{Bunster:2014mua} C.~Bunster, M.~Henneaux, A.~Perez, D.~Tempo
and R.~Troncoso, ``Generalized Black Holes in Three-dimensional
Spacetime,'' JHEP \textbf{1405}, 031 (2014) {[}arXiv:1404.3305 {[}hep-th{]}{]}.
%doi:10.1007/JHEP05(2014)031      %%CITATION = doi:10.1007/JHEP05(2014)031;%%

\bibitem{BTZ}M.~Banados, C.~Teitelboim and J.~Zanelli,   ``The Black hole in three-dimensional space-time,''   Phys.\ Rev.\ Lett.\  {\bf 69}, 1849 (1992)   doi:10.1103/PhysRevLett.69.1849   [hep-th/9204099].   %%CITATION = doi:10.1103/PhysRevLett.69.1849;%%

\bibitem{BHTZ}M.~Banados, M.~Henneaux, C.~Teitelboim and J.~Zanelli,   ``Geometry of the (2+1) black hole,''   Phys.\ Rev.\ D {\bf 48}, 1506 (1993)   Erratum: [Phys.\ Rev.\ D {\bf 88}, 069902 (2013)]   doi:10.1103/PhysRevD.48.1506, 10.1103/PhysRevD.88.069902   [gr-qc/9302012].   %%CITATION = doi:10.1103/PhysRevD.48.1506, 10.1103/PhysRevD.88.069902;%%

\bibitem{MarikaTaylor-non-relativistic} M.~Taylor,   ``Non-relativistic holography,''   arXiv:0812.0530 [hep-th].   %%CITATION = ARXIV:0812.0530;%%

\bibitem{Bertoldi:2009vn}G.~Bertoldi, B.~A.~Burrington and A.~Peet,   ``Black Holes in asymptotically Lifshitz spacetimes with arbitrary critical exponent,''   Phys.\ Rev.\ D {\bf 80}, 126003 (2009)   doi:10.1103/PhysRevD.80.126003   [arXiv:0905.3183 [hep-th]].   %%CITATION = doi:10.1103/PhysRevD.80.126003;%%

\bibitem{Bertoldi:2009dt}G.~Bertoldi, B.~A.~Burrington and A.~W.~Peet,   ``Thermodynamics of black branes in asymptotically Lifshitz spacetimes,''   Phys.\ Rev.\ D {\bf 80}, 126004 (2009)   doi:10.1103/PhysRevD.80.126004   [arXiv:0907.4755 [hep-th]].   %%CITATION = doi:10.1103/PhysRevD.80.126004;%%

\bibitem{DHoker-Kraus}E.~D'Hoker and P.~Kraus,   ``Holographic Metamagnetism, Quantum Criticality, and Crossover Behavior,''   JHEP {\bf 1005}, 083 (2010)   doi:10.1007/JHEP05(2010)083   [arXiv:1003.1302 [hep-th]].   %%CITATION = doi:10.1007/JHEP05(2010)083;%%

\bibitem{Hartnoll} S.~A.~Hartnoll, D.~M.~Ramirez and J.~E.~Santos,   ``Emergent scale invariance of disordered horizons,''   JHEP {\bf 1509}, 160 (2015)   doi:10.1007/JHEP09(2015)160   [arXiv:1504.03324 [hep-th]].   %%CITATION = doi:10.1007/JHEP09(2015)160;%%

\bibitem{Marika-Taylor-Lifshitz-holo}M.~Taylor,   ``Lifshitz holography,''   Class.\ Quant.\ Grav.\  {\bf 33}, no. 3, 033001 (2016)   doi:10.1088/0264-9381/33/3/033001   [arXiv:1512.03554 [hep-th]].   %%CITATION = doi:10.1088/0264-9381/33/3/033001;%%

\bibitem{HernanTT}H.~A.~Gonzalez, D.~Tempo and R.~Troncoso,   ``Field theories with anisotropic scaling in 2D, solitons and the microscopic entropy of asymptotically Lifshitz black holes,''   JHEP {\bf 1111}, 066 (2011)   doi:10.1007/JHEP11(2011)066   [arXiv:1107.3647 [hep-th]].   %%CITATION = doi:10.1007/JHEP11(2011)066;%%

\bibitem{Cardy} J.~L.~Cardy,   ``Operator Content of Two-Dimensional Conformally Invariant Theories,''   Nucl.\ Phys.\ B {\bf 270}, 186 (1986).   doi:10.1016/0550-3213(86)90552-3   %%CITATION = doi:10.1016/0550-3213(86)90552-3;%%

\bibitem{DiFrabcescolibro}P.~Di Francesco, P.~Mathieu and D.~Senechal,   ``Conformal Field Theory,''   doi:10.1007/978-1-4612-2256-9   %%CITATION = doi:10.1007/978-1-4612-2256-9;%%

\bibitem{Carlip:1999cy}S.~Carlip,   ``Entropy from conformal field theory at Killing horizons,''   Class.\ Quant.\ Grav.\  {\bf 16}, 3327 (1999)   doi:10.1088/0264-9381/16/10/322   [gr-qc/9906126].   %%CITATION = doi:10.1088/0264-9381/16/10/322;%%

\bibitem{Correa-Troncoso-martinez1}F.~Correa, C.~Martinez and R.~Troncoso,   ``Scalar solitons and the microscopic entropy of hairy black holes in three dimensions,''   JHEP {\bf 1101}, 034 (2011)   doi:10.1007/JHEP01(2011)034   [arXiv:1010.1259 [hep-th]].   %%CITATION = doi:10.1007/JHEP01(2011)034;%%

\bibitem{Correa-Troncoso-Martinez2}F.~Correa, C.~Martinez and R.~Troncoso,   ``Hairy Black Hole Entropy and the Role of Solitons in Three Dimensions,''   JHEP {\bf 1202}, 136 (2012)   doi:10.1007/JHEP02(2012)136   [arXiv:1112.6198 [hep-th]].   %%CITATION = doi:10.1007/JHEP02(2012)136;%%

\bibitem{Dauxois:2006zz}T.~Dauxois and M.~Peyrard,  ``Physics of solitons,''Cambridge, UK: Univ. Pr. (2006) 422 p.

\bibitem{Daniel} H.~Afshar, S.~Detournay, D.~Grumiller, W.~Merbis, A.~Perez, D.~Tempo and R.~Troncoso,   ``Soft Heisenberg hair on black holes in three dimensions,''   arXiv:1603.04824 [hep-th].   %%CITATION = ARXIV:1603.04824;%%

\bibitem{Hawking}S.~W.~Hawking, M.~J.~Perry and A.~Strominger,   ``Soft Hair on Black Holes,''   arXiv:1601.00921 [hep-th].   %%CITATION = ARXIV:1601.00921;%% 

\bibitem{Hartnoll:2009sz}S.~A.~Hartnoll,   ``Lectures on holographic methods for condensed matter physics,''   Class.\ Quant.\ Grav.\  {\bf 26}, 224002 (2009)   doi:10.1088/0264-9381/26/22/224002   [arXiv:0903.3246 [hep-th]].   %%CITATION = doi:10.1088/0264-9381/26/22/224002;%%

\bibitem{Hartnoll:2011fn}S.~A.~Hartnoll,   ``Horizons, holography and condensed matter,''   arXiv:1106.4324 [hep-th].   %%CITATION = ARXIV:1106.4324;%%

\bibitem{Eloy}E.~Ayón-Beato, M.~Bravo-Gaete, F.~Correa, M.~Hassaïne, M.~M.~Juárez-Aubry and J.~Oliva,   ``First law and anisotropic Cardy formula for three-dimensional Lifshitz black holes,''   Phys.\ Rev.\ D {\bf 91}, no. 6, 064006 (2015)   doi:10.1103/PhysRevD.91.064006   [arXiv:1501.01244 [gr-qc]].   %%CITATION = doi:10.1103/PhysRevD.91.064006;%%

\bibitem{Hassaine1}M.~Bravo-Gaete, S.~Gomez and M.~Hassaine,   ``Cardy formula for charged black holes with anisotropic scaling,''   Phys.\ Rev.\ D {\bf 92}, no. 12, 124002 (2015)   doi:10.1103/PhysRevD.92.124002   [arXiv:1510.04084 [hep-th]].   %%CITATION = doi:10.1103/PhysRevD.92.124002;%%

\bibitem{BarnichFlat}G.~Barnich,   ``Entropy of three-dimensional asymptotically flat cosmological solutions,''   JHEP {\bf 1210}, 095 (2012)   doi:10.1007/JHEP10(2012)095   [arXiv:1208.4371 [hep-th]].   %%CITATION = doi:10.1007/JHEP10(2012)095;%%

\bibitem{Bagchi-Detournay}A.~Bagchi, S.~Detournay, R.~Fareghbal and J.~Simón,   ``Holography of 3D Flat Cosmological Horizons,''   Phys.\ Rev.\ Lett.\  {\bf 110}, no. 14, 141302 (2013)   doi:10.1103/PhysRevLett.110.141302   [arXiv:1208.4372 [hep-th]].   %%CITATION = doi:10.1103/PhysRevLett.110.141302;%% 

\bibitem{Detournay-Hartman}S.~Detournay, T.~Hartman and D.~M.~Hofman,   ``Warped Conformal Field Theory,''   Phys.\ Rev.\ D {\bf 86}, 124018 (2012)   doi:10.1103/PhysRevD.86.124018   [arXiv:1210.0539 [hep-th]].   %%CITATION = doi:10.1103/PhysRevD.86.124018;%%

\bibitem{shag}E.~Shaghoulian,   ``A Cardy formula for holographic hyperscaling-violating theories,''   JHEP {\bf 1511}, 081 (2015)   doi:10.1007/JHEP11(2015)081   [arXiv:1504.02094 [hep-th]].   %%CITATION = doi:10.1007/JHEP11(2015)081;%%

\bibitem{Hassaine3}M.~Bravo-Gaete, S.~Gomez and M.~Hassaine,   ``Towards the Cardy formula for hyperscaling violation black holes,''   Phys.\ Rev.\ D {\bf 91}, no. 12, 124038 (2015)   doi:10.1103/PhysRevD.91.124038   [arXiv:1505.00702 [hep-th]].   %%CITATION = doi:10.1103/PhysRevD.91.124038;%%

\bibitem{Castro-Hoffman}A.~Castro, D.~M.~Hofman and G.~Sárosi,   ``Warped Weyl fermion partition functions,''   JHEP {\bf 1511}, 129 (2015)   doi:10.1007/JHEP11(2015)129   [arXiv:1508.06302 [hep-th]].   %%CITATION = doi:10.1007/JHEP11(2015)129;%%

\bibitem{Afshar-Detournay}H.~Afshar, S.~Detournay, D.~Grumiller and B.~Oblak,   ``Near-Horizon Geometry and Warped Conformal Symmetry,''   JHEP {\bf 1603}, 187 (2016)   doi:10.1007/JHEP03(2016)187   [arXiv:1512.08233 [hep-th]].   %%CITATION = doi:10.1007/JHEP03(2016)187;%%

\bibitem{Detournay-Celine}S.~Detournay, L.~A.~Douxchamps, G.~S.~Ng and C.~Zwikel, ``Warped AdS$_3$ Black Holes in Higher Derivative Gravity Theories,''   arXiv:1602.09089 [hep-th].   %%CITATION = ARXIV:1602.09089;%%

\bibitem{Abdalla}E.~Abdalla, J.~de Oliveira, A.~Lima-Santos and A.~B.~Pavan,   ``Three dimensional Lifshitz black hole and the Korteweg-de Vries equation,''   Phys.\ Lett.\ B {\bf 709}, 276 (2012)   doi:10.1016/j.physletb.2012.02.026   [arXiv:1108.6283 [hep-th]].   %%CITATION = doi:10.1016/j.physletb.2012.02.026;%%

\bibitem{Carlip-Teitelboim} S.~Carlip and C.~Teitelboim,   ``Aspects of black hole quantum mechanics and thermodynamics in (2+1)-dimensions,''   Phys.\ Rev.\ D {\bf 51}, 622 (1995)   doi:10.1103/PhysRevD.51.622   [gr-qc/9405070].   %%CITATION = doi:10.1103/PhysRevD.51.622;%%

\bibitem{Maldacena-Strominger} J.~M.~Maldacena and A.~Strominger,   ``AdS(3) black holes and a stringy exclusion principle,''   JHEP {\bf 9812}, 005 (1998)   doi:10.1088/1126-6708/1998/12/005   [hep-th/9804085].   %%CITATION = doi:10.1088/1126-6708/1998/12/005;%%

\bibitem{FracKdV1}Johnson, R. S. (1970). A non-linear equation incorporating damping and dispersion. Journal of Fluid Mechanics, 42(01), 49-60.

\bibitem{FracKdV2}Feng, Z. (2002). On explicit exact solutions to the compound Burgers-KdV equation. Physics Letters A, 293(1), 57-66.

\bibitem{FracKdV3}Wang, Q. (2006). Numerical solutions for fractional KdV-Burgers equation by Adomian decomposition method. Applied Mathematics and Computation, 182(2), 1048-1055.

\bibitem{FracKdV4}Younis, M. (2014). Soliton Solutions of Fractional Order KdV-Burger's Equation. Journal of Advanced Physics, 3(4), 325-328.

\bibitem{Klebanov:1999tb}I.~R.~Klebanov and E.~Witten,   ``AdS / CFT correspondence and symmetry breaking,''   Nucl.\ Phys.\ B {\bf 556}, 89 (1999)   doi:10.1016/S0550-3213(99)00387-9   [hep-th/9905104].

\bibitem{Witten:2001ua}E.~Witten,   
``Multitrace operators, boundary conditions, and AdS / CFT correspondence,''
hep-th/0112258.   %%CITATION = HEP-TH/0112258;%%   %306 citations counted in INSPIRE as of 03 May 2016

\bibitem{Shomer}A.~Sever and A.~Shomer,   ``A Note on multitrace deformations and AdS/CFT,''   JHEP {\bf 0207}, 027 (2002)   doi:10.1088/1126-6708/2002/07/027   [hep-th/0203168].   %%CITATION = doi:10.1088/1126-6708/2002/07/027;%%

\bibitem{de Boer}J.~de Boer and D.~Engelhardt,   ``Comments on Thermalization in 2D CFT,''   arXiv:1604.05327 [hep-th].   %%CITATION = ARXIV:1604.05327;%% 

\bibitem{Compere-Song}G.~Compère and W.~Song,   ``$\mathcal{W}$ symmetry and integrability of higher spin black holes,''   JHEP {\bf 1309}, 144 (2013)   doi:10.1007/JHEP09(2013)144   [arXiv:1306.0014 [hep-th]].   %%CITATION = doi:10.1007/JHEP09(2013)144;%%

\bibitem{Gutperle1}M.~Gutperle and Y.~Li,   ``Higher Spin Lifshitz Theory and Integrable Systems,''   Phys.\ Rev.\ D {\bf 91}, no. 4, 046012 (2015)   doi:10.1103/PhysRevD.91.046012   [arXiv:1412.7085 [hep-th]].   %%CITATION = doi:10.1103/PhysRevD.91.046012;%%

\bibitem{Gutperle2}M.~Beccaria, M.~Gutperle, Y.~Li and G.~Macorini,   ``Higher spin Lifshitz theories and the Korteweg-de Vries hierarchy,''   Phys.\ Rev.\ D {\bf 92}, no.8,085005 (2015)   doi:10.1103/PhysRevD.92.085005   [arXiv:1504.06555 [hep-th]].   %%CITATION = doi:10.1103/PhysRevD.92.085005;%%

\bibitem{Drinfeld:1984qv}V.~G.~Drinfeld and V.~V.~Sokolov,   ``Lie algebras and equations of Korteweg-de Vries type,''   J.\ Sov.\ Math.\  {\bf 30}, 1975 (1984).   doi:10.1007/BF02105860   %%CITATION = doi:10.1007/BF02105860;%%

\bibitem{Mathieu:1991et}P.~Mathieu and W.~Oevel,   ``The W(3)(2) conformal algebra and the Boussinesq hierarchy,''   Mod.\ Phys.\ Lett.\ A {\bf 6}, 2397 (1991).   doi:10.1142/S0217732391002827   %%CITATION = doi:10.1142/S0217732391002827;%%

\bibitem{Das:1990td}A.~K.~Das, W.~J.~Huang and S.~Roy,   ``The Zero Curvature Formulation Of The Boussinesq Equation,''   Phys.\ Lett.\ A {\bf 163}, 186 (1991).   doi:10.1016/0375-9601(91)90791-6
\end{thebibliography}
\end{document}